# Metrology Camera System of Prime Focus Spectrograph for Subaru Telescope

Shiang-Yu Wang[a], Richard C.-Y. Chou[a], Yin-Chang Chang[a], Pin-Jie Huang[a], Yen-Sang Hu[a], Hsin-Yo Chen[a], Naoyuki Tamura[b], Naruhisa Takato[c], Hung-Hsu Ling[a], James E. Gunn[d], Jennifer Karr[a], Chi-Hung Yan[a], Peter Mao[e], Youichi Ohyama[a], Hiroshi Karoji[b], Hajime Sugai[b], Atsushi Shimono[b]

[a]Academia Sinica, Institute of Astronomy and Astrophysics, P. O. Box 23-141, Taipei, Taiwan;
[b]Kavli Institute for the Physics and Mathematics of the Universe (WPI), the University of Tokyo, 5-1-5 Kashiwanoha, Kashiwa, 277-8583, Japan;
[c]Subaru Telescope, National Astronomical Observatory of Japan, 650 North A'ohoku Place Hilo, HI 96720, U.S.A.
[d]Princeton University, Princeton, New Jersey, 08544, USA
[e]California Institute of Technology, 1200 E California Blvd, Pasadena, CA 91125, USA.

## ABSTRACT

The Prime Focus Spectrograph (PFS) is a new optical/near-infrared multi-fiber spectrograph designed for the prime focus of the 8.2m Subaru telescope. The metrology camera system of PFS serves as the optical encoder of the COBRA fiber motors for the configuring of fibers. The 380mm diameter aperture metrology camera will locate at the Cassegrain focus of Subaru telescope to cover the whole focal plane with one 50M pixel Canon CMOS sensor. The metrology camera is designed to provide the fiber position information within 5µm error over the 45cm focal plane. The positions of all fibers can be obtained within 1s after the exposure is finished. This enables the overall fiber configuration to be less than 2 minutes.

## 1. INTRODUCTION

The Prime Focus Spectrograph (PFS) is a new prime focus optical/near-infrared multi-fiber spectrograph of the 8.2m Subaru telescope[1]. PFS will cover 1.3 degree diameter field with 2394 fibers to complement the imaging capability of Hyper SuprimeCam[2]. To retain high throughput of PFS, the final positioning accuracy between the fibers and observing targets is required to be less than 10µm. The metrology camera serves as the optical encoder of the fiber motors for the configuring of fibers[3]. The metrology camera is designed to provide the fiber position information within 5µm error over the 45cm focal plane. The information from the metrology camera will be fed into the fiber positioner control system for the close loop control.

The metrology camera will locate at the Cassegrain focus of Subaru telescope to cover the whole focal plan with one 50M pixel Canon CMOS sensor. To reduce the high spatial frequency distortion of the wide field corrector (WFC), the aperture size of the metrology camera is set to be 380mm which is the largest allowed aperture at the Cassegrain focus. A Schmidt telescope type optical design was adapted to provide uniform image quality across the field with reasonable sampling of the point spreading function. The mechanical design based on Invar provides stable structure over the temperature range under operation conditions. The CMOS sensor can be read in 0.8s to reduce the overhead for the fiber configuration. The positions of all fibers can be obtained within 1s after the exposure is finished. This enables the overall fiber configuration to be less than 2 minutes.

# 2. METROLOGY CAMERA DESIGN

The metrology camera will be installed inside a Cassegrain instrument box like other Subaru Cassegrain instruments. Calibration of the image distortion and mapping to the focal plane through the WFC is achieved by back-lit fixed fiducial fibers and the home positions of scientific fibers. The home position of the science fibers have been tested to provide good repeatability (~1µm) after rotational movements. The positions for the fiducial fibers and the home positions of science fibers will be measured to a high repeatable accuracy during the integration and verification of the fiber system.

In the beginning of each COBRA configuration, COBRAs will move to the home position to generate the reference points across the field for the distortion map estimation. After that, the science fibers will move to new positions based on the model derived from the chosen target field. The metrology camera system will measure the location of each fiber position and send it to the COBRA control system. Then, the fiber position errors are calculated with respect to the required locations for the fibers to move to the new positions. This iteration will be repeated for a number of times until the errors are smaller than the expected value (10µm). The back illuminator of fiducial and science fibers will be lit up accordingly through the iterations. In addition, the metrology camera will also support the other two diagnostic functions during the commissioning and engineering time. The first is to image the circular motion of the COBRA fibers to get the rotational center of the motors. During such operation, the back illuminated source will be triggered by 10ms pulses with duty cycle of 10% to generate spots on the movement circle. Metrology camera system will calculate the center positions of the circles with high precision. The second function is to generate the fiber images when the instrument rotator is moving without any COBRA move. With few seconds of exposure with back illuminator on, the metrology camera will image the arcs generated by the fibers and estimate the offset and tilt between the rotator axis center to the center of the fibers focal plane.

The details of the components of the metrology camera are described in the following sections.

## 2.1 Camera optics

The key requirement for the optics of the metrology camera is to provide uniform optical performance and negligible image distortion across the field. This ensures the well-controlled centroid estimation error of the metrology camera for all fibers. The optics of metrology camera produces an image of the prime focal plane with required magnification factor. It should be noticed that the centroid estimation accuracy is better with higher magnification factor. Considering the sensitive area size of the detector used, a magnification factor of 0.038 was chosen for the metrology camera.

The determination of the metrology camera aperture size has been changed due to the characteristics of the WFC. Unlike the target star light path of PFS, the metrology camera locating at the Cassegrain focus observes the light from fibers at the prime focal plane. Since the fiber tips are very close to the first WFC lens surface, the footprint of fibers on the WFC lens surfaces is quite small (smaller than 1mm) if the aperture size of the metrology camera is not big enough. Under such condition, the light beam is deviated by the high spatial frequency manufacturing errors on lens surfaces. According to the report from Canon who fabricated WFC, the local peak to valley surface error of WFC is about 10 - 30 nm with 6 mm period due to the polishing tool size. Figure 1 presents the standard deviation of the beam deviation as a function of the metrology camera aperture size. This local surface error will generate about 6 µm of beam deviation for a 110mm aperture metrology camera which was design in the earlier phase of the project[3]. It is larger than the 5µm centroid estimation error assigned for the metrology camera.

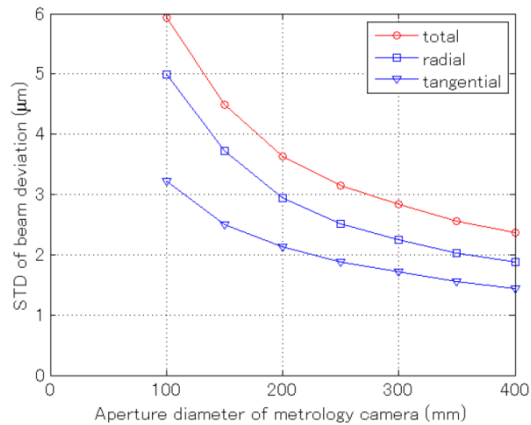

Figure 1. Standard deviation of beam deviation versus metrology camera aperture size. For a 100 mm camera, 6 µm of beam deviation is predicted, which is obviously larger than the required position accuracy of 5µm.

As shown in Figure 1, the aperture size of the metrology camera should be as large as possible to minimize the possible error from the WFC. Considering the central hole size at the Cassegrain focus of Subaru telescope, the aperture size of the metrology camera is set to be 380mm. With a 380mm aperture, the beam deviation can be reduced to be around 2.5µm.

With the 380mm aperture size and the magnification factor of 0.038, the metrology camera requires a fast optics. Furthermore, the Cassegrain box has a height limit about 1750mm. Thus, we chose the Schmidt reflector design which consists of a Schmidt plate, a spherical mirror and a field flattener for the metrology camera. Figure 2 shows the optical layout of the design. The focal length is about 840 mm and the focal ratio of f/2.2. Compared with the 110mm aperture design, it is no longer telecentric and has a telecentric angle of ~ 1.0 degree at the field edge. In order to generate well sampled image spots, the Schmidt plate was deliberately designed to under correct the spherical aberration. The predicted spot size of the fiber tip extends about 10µm or ~3 pixels. The shape of the point spread functions all over the entire field is quite uniform and satisfies the centroid measurement requirement. Figure 3 shows the simulated image at the filed center and edge with the possible manufacturing and alignment errors. The distortion at the field edge is 0.0001 which is much smaller than the distortion of WFC (0.0097).

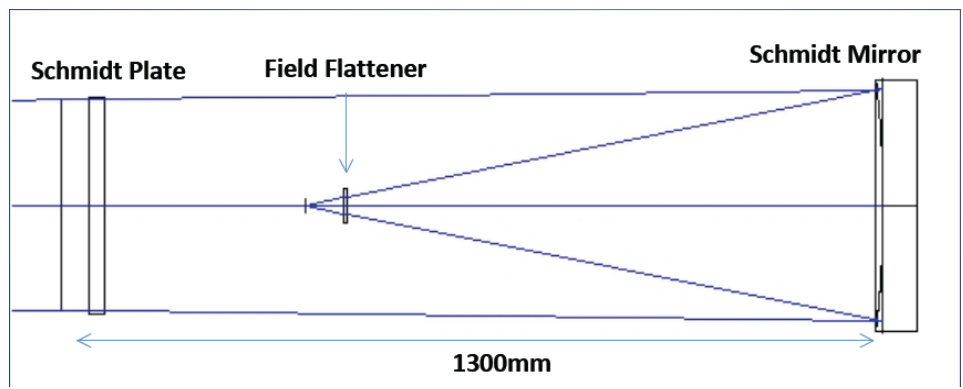

Figure 2. The optical layout of the metrology camera.

The possible venders for the mirror and the lenses have been identified. All the lenses are made by silica and coated with narrow pass band (30nm FWHM) coating to fit the back lit light source and provide the possibility of operating the metrology camera at day time for system calibration.

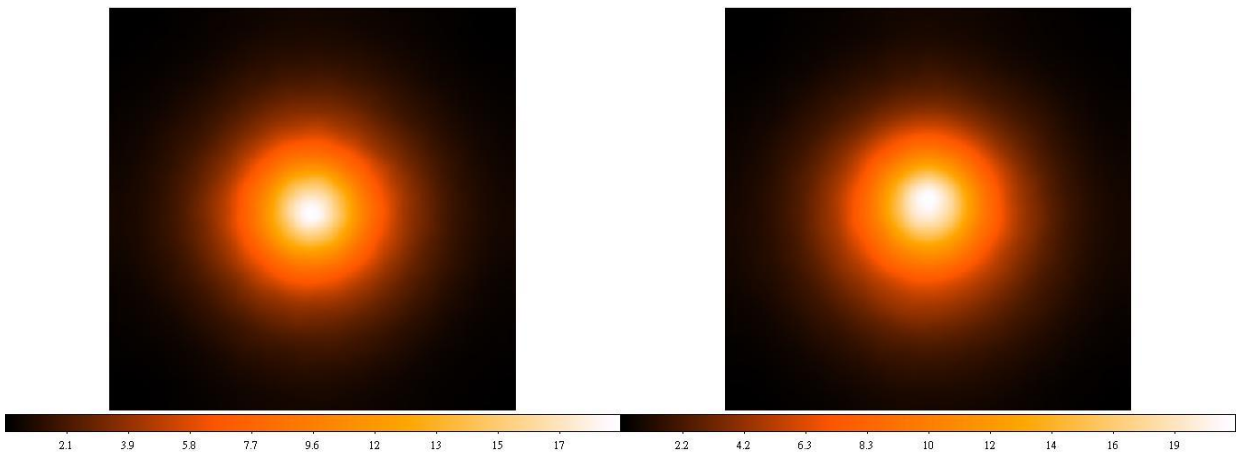

Figure 3. The point spreading function for field center (left) and field edge (right) of the PFS metrology camera. The point spread functions have a FWHM of ~ 10µm across the entire field and satisfy the centroid measurement requirement.

## 2.2 Camera detector

The camera sensor used is a 50M pixel CMOS sensor with 8960 × 5778 pixels and 3.2μm pixel size developed by Canon Inc, Figure 4 is the picture for the sensor. This large format sensor can cover the entire focal plane and provide the required centroid estimation accuracy of fibers with a single exposure. The physical size of the sensor is about 28.67mm × 18.49mm. It is a frontside illuminated sensor with a micro-lens on every pixel to improve the light collection efficiency. With the current optical design, each fiber image will be sampled by roughly 3 pixels for the 10μm FWHM spots.

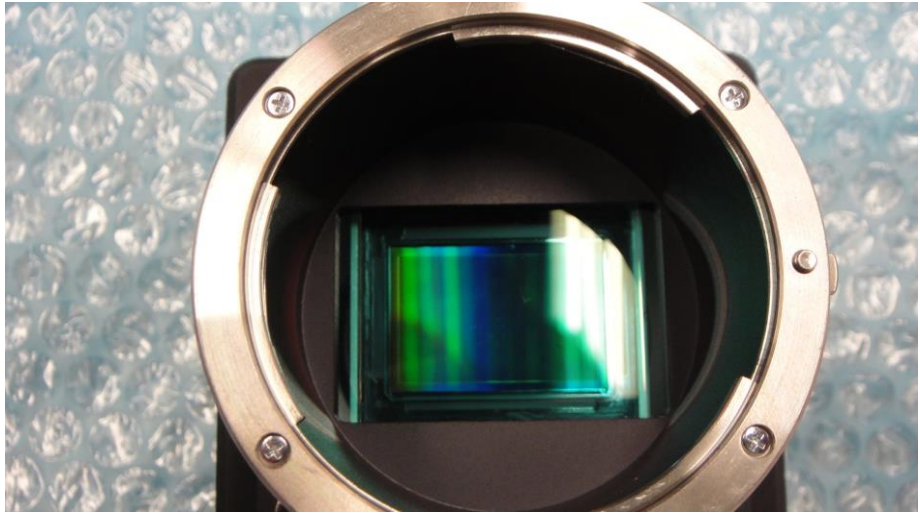

Figure 4. The Canon 50M pixel CMOS sensor.

The sensor has 8 interlaced output channels and the adjacent pixels are designed to be read from different outputs. It is a 12 bit system with four different gain settings. The camera provides the rolling shutter operation only. The performance of the camera was tested with the standard photon transfer curve method. The conversion gains for the different settings are 2.24, 1.16, 0.69 and 0.39 $e^-$/DN respectively. The readout noise decreases with the conversion gain. The readout noise for the four different settings is 5.34, 3.67, 2.74 and 2.06 $e^-$ respectively from high to low gain settings. Although the lowest gain mode gives the lowest readout noise, the highest signal to noise ratio is obtained with the highest gain mode. Since the fiber illuminator intensity can be adjusted by design, we will use the highest gain mode for the metrology camera and it can provide the largest dynamical range (~ 2000).

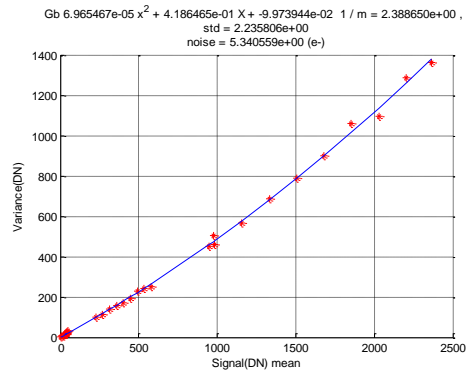

Figure 5. The photon transfer curve for the highest gain setting of Canon CMOS sensor.

There is no temperature regulation for the camera so the dark measurement might vary with sensor temperature. However, no dark current was detected at the environment temperature between 20~25 degree Celsius with 1s exposure time. Given the exposure time allowed for the metrology camera and the lower operation temperatures on Mauna Kea, we do not expect the dark current to contribute excess noise. The peak quantum efficiency is around 50% at 520nm as shown in Figure 6. The linearity of different gain settings was also checked. The non-linearity is smaller than 1% of the full 12-bit resolution. The characteristics of the sensor are very close to the scientific CCDs despite of a lower QE. Since the metrology camera receives the signal from the light source through fibers, lower QE is not a critical parameter for our application.

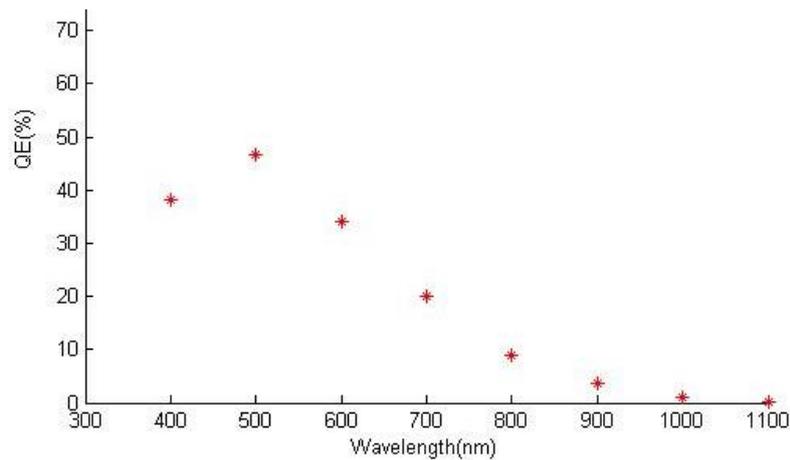

Figure 6. The measured quantum efficiency of the CMOS sensor.

## 2.3 Camera mechanical design

To be treated as a Subaru Cassegrain instrument, the mechanical structure for the metrology camera should meet the space and mounting scheme like other Cass instruments. The height of the Cass box is 1750mm which gives a strong constrain to the optical design of the metrology camera. The mechanical structure for the metrology camera can thus be separated into two components: 1) the supporting and mounting structure for the optical elements and camera sensor module; 2) a box with mounting flange to match the telescope interface. This provides the required interface to fit our system to Cassgrain bonette of Subaru telescope so that the system can be positioned with a repeatable accuracy. The camera mounting structure is connected to the Cassegrain box through a top plate. Figure 7 shows the design of supporting structure and the Caasegrain box of the metrology camera. The optical elements are mounted with eight supporting beams. The Schmidt plate is very close to the fixed top plate and it is mounted directly to it. The field flattener and the sensor module are mounted together with a high precision focusing stage to adjust the focal plane position for optical performance compensation. The spherical mirror will be supported from the side with kinetic mounts on the bottom to hold the mirror at the correct axial position. The spherical mirror and the field flattener with initial alignment capability are required to achieve the expected performance during the commissioning.

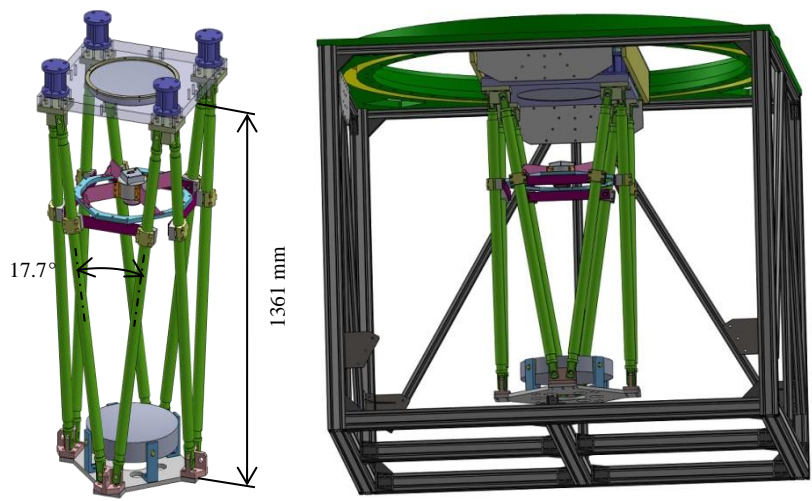

Figure 7. The mechanical supporting structure of the metrology camera (left) and the Cassegrain box structure (right).

The thermal and structure analysis for the supporting structure was made with Invar as well as Aluminum 6061 alloy to better understand the optical performance change under different observing conditions. For thermal expansion analysis, the temperature variation was set from -5 to 5 degrees Celsius to meet the environment conditions on Mauna Kea. For the invar structure, the largest thermal expansion/contraction is on the order of one µm and can be ignored. For the aluminum structure, the analysis shows that the focal plane needs to shift +/- 70µm for +/-5 degrees Celsius to compensate the thermal expansion/contraction.

For structure deformation analysis, the deformation was derived with ANSYS and added into the ZEMAX design to simulate the optical performance change under different elevation angles. With the elongate structure and heavy weight mirror at the bottom of the structure, the shift of the mirror center is about 200µm for both structures when the telescope moves from Zenith to 15 degree from horizon. Such deformation definitely changes the focal position of the metrology camera. To evaluate the image quality and the focal plane position error, the focal plane was shifted from -100µm to 100µm away from the focus position at Zenith in the Zemax simulation with 10µm step. Then, we derived the centroid accuracy using point spread functions obtained from different focal plane offsets. The results are shown in Figure 8. Both supporting structures show similar centroid accuracy as a function of focal plane offset because the derived structure deformations for two materials are more or less the same. Given the acceptable centroid accuracy should be less than 0.02 pixels (dotted lines in Figure 8) the acceptable focal plane offset should be ~ +/- 25µm away from the optimized location. Refocusing is needed when the telescope elevation angle is changed for more than 30 degrees. We will choose Invar for the supporting structure although both materials give similar tilt deformation results. The invar structure is almost free from the thermal expansion/contraction due to temperature variation and therefore can deliver a more stable performance.

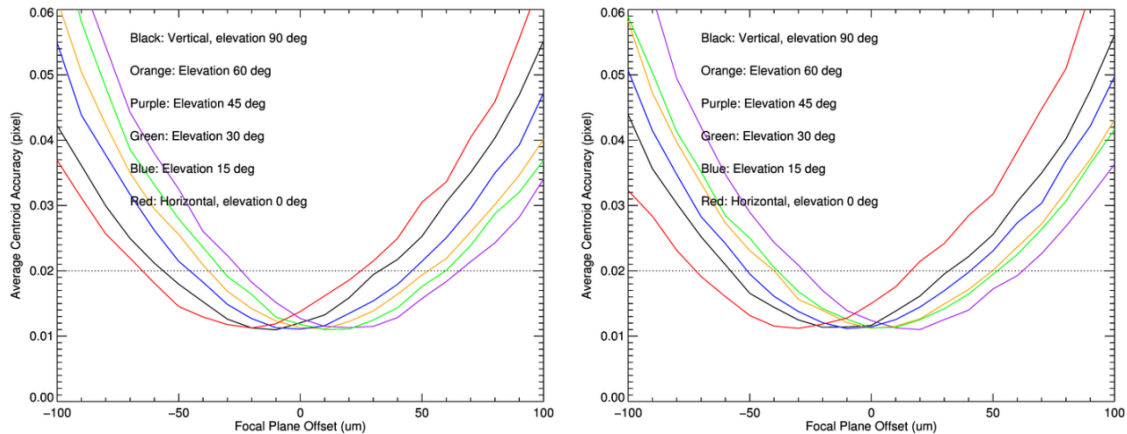

Figure 8. Average centroid accuracy over the entire field of the metrology camera with respect to the focal plane offset under various elevation angles. The result for the Invar (Aluminum) structure is shown on the right (left) panel. The centroid accuracy for both results are very similar because the deformation of the two supporting structure is more or less the same. Although aluminum is less strong compared with invar, the light weight of aluminum mitigates the structure deformation. The horizontal dotted line represents the threshold for the acceptable centroid accuracy. This suggests that the acceptable focal plane offset for both structures should be ~ +/- 25µm away from the optimized location.

**2.4 Control and operation**

The readout and configuration control of the 50M pixel camera is based on the API provided by Canon with the high level scripts developed in house. The CMOS sensor is linked to the control computer with CameraLink interface. The readout time for the 50M pixel camera is 0.8 second. The raw image data will be stored in the memory of the control computer to allow fast access and image process for the centroid calculation.

The main task of the metrology camera software is to identify the science and fiducial fibers and measure the centroids of fiber spots. These centroids along with fiber identifications are then sent to the COBRA control system. The image will be reduced by the software including pixel sensitivity adjustment, pixel DC offset, and excluding bad pixels. The camera software receives system commands and information and sends the measured fiber centroid positions to the PFS control system. The PFS control system also controls the back illumination sources so the fibers are lit on at the right time for the metrology camera. In addition, the metrology camera software provides diagnostic and test capabilities of the CMOS detectors.

The centroid calculation algorithm is very important to provide solid centroid calculation in a short time for all fibers. Based on our tests with FMOS fiber images, we have developed a modified DAOPHOT algorithm for the centroid estimation. The algorithm locates fiber positions by checking through all pixels. A threshold of pixel value is given to ignore the low SNR spot generated by the noise. Then the convolution of the image with a Gaussian profile is executed to find the centroid of the fiber images. This process will get the locations of the local maximum of the image. The next step is to filter out the sources with other criteria, such as spot sharpness and roundness to remove hot pixels, cosmic rays or other unwanted fake spots.

To verify the centroid algorithm can provide the centroid for all fibers in a short time, the tests have been run on simulated data matching the instrument in terms of the image dimension and number of fibers. Figure 9 shows the time required to get the centroid of the spot for different size of image and different number of spots. This test was done by a Python version of DAOPHOT. The processing time is linear with number of pixels. The biggest computational expense in the psf finding routine is the convolution, which scales as the number of pixels. The number of fibers produces a minor secondary effect at the values we are dealing with. In all cases, the processing time is nearly constant until around 5000 fibers. The algorithm starts to slow down with more than 10000 spots in the images.

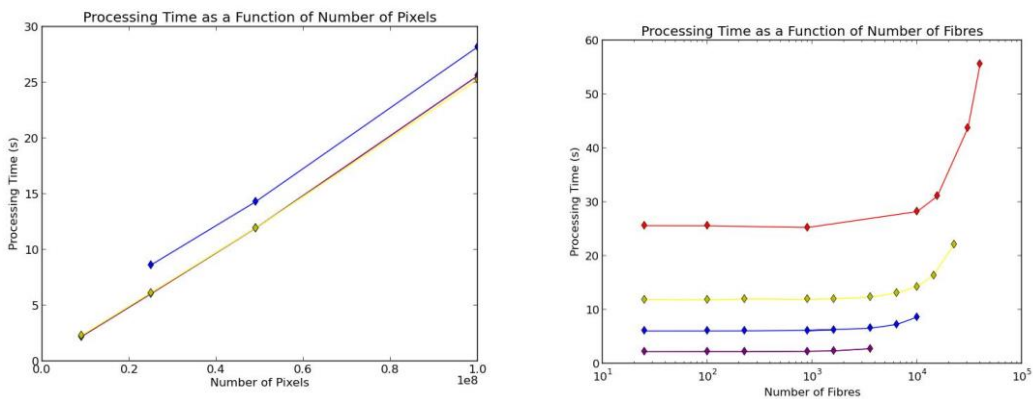

Figure 9. The image processing time for different image size (left) and fiber numbers in the image(right). Three different numbers of fibers (25 (yellow) 900 (purple) and 10000 (blue)) were shown in the left figure. Four different image sizes (3000x3000 (purple), 5000x5000 (blue), 7000x7000 (yellow) and 10000x10000 (red)) were shown in the right figure.

With the pixel numbers and the fiber numbers of the PFS design, the computational time for each steps of the centroid algorithm was examined. Over 92% of the time in the computation is for the convolution. The second largest expense is reading in the FITS file. Both processes can be parallelized in a straightforward fashion using threads, by performing the algorithm on subsets of the image. The individual threads do not share resources, which simplifies the parallelization. We have generated a C program of modified DAOPHOT process and tested the parallel process. Initial tests perform as expected, with four cores instead of one reducing the running time to 28% of the original time (i.e. almost four times faster). The current time to process a single image on a four core machine is 10s. The current convolution routine and the code could be made more efficient by converting the 2D convolution into 1D convolution for a 4.5 times increase in speed. Factoring in the convolution routine, this should decrease to 2.2s. With the expected increase in efficiency produced by this, the point source detection of the image will easily be doable on a 16 core machine for a processing time about 0.6s.

## 3. METROLOGY SYSTEM ANALYSIS

The most critical requirement for the metrology camera is to determine the positions of the fibers on the focal plane within an error of 5µm. The centroid estimation error of the fibers at the focal plane comes from the following factors

1. The image center estimation error: This includes the optical quality of the fiber image, the point spreading function sampling and the centroid estimation algorithm. As estimated from the simulated images discussed in section 2.1, the error is controlled to be smaller than 0.02 pixels which correspond to 1.7µm at the PFS focal plane.

2. The WFC surface figure error: As shown in Figure 1, the possible error from the WFC with 380mm aperture camera is about 2.5 µm. However, we expect such error can be reduced when PFS start some observations at the telescope. With the raster scan we plan for the commissioning process and increase of the different fiber configurations, we will be able to estimate the WFC surface error. Since the WFC surface error is fixed instead of the variable parameter, we will be able to estimate it and reduce its contribution to the centroid estimation error.

3. The dome seeing effect: The distance between the metrology camera and the PFS focal plane is about 18 meters. The light path will suffer from the possible turbulence inside the dome. The light passing through the turbulence will deviate from its original position on the camera sensor. In order to understand such effect, a series of tests were executed with Subaru FMOS[4] instrument under different conditions. The details are discussed in the following section:

### 3.1 Dome seeing effect

The FMOS fiber tips were imaged through WFC for FMOS and the telescope lens of a commercial CCD camera mounted on the Cassegain focus with camera pixel size of 3.45μm. The fibers were in home position. The magnification factor of the camera is about 0.0706. The FMOS fiber size is 100μm so it is about 7μm on the CCD plane. The CCD readout time is 6s for the whole frame. For most data, two rows of FMOS fibers were illuminated and only 2400×300 pixels of the CCD were read for the images to minimize the readout time. The FoV of the camera roughly covers 115mm of the FMOS focal plane. The image spot size FWHM is around 9μm which is close the diffraction limit of 8.75μm. No filter is added in the optical path. Series of exposures was taken on Dec 24, 25 and 26 in 2012. In each series, exposures of 0.1, 0.3, 0.5, 1, 2 and 5 seconds with 100 frames were taken. The same series of exposures were repeated at telescope elevation angle of 30, 60 and 90 degrees under both dome open and dome closed conditions. The SNR for the images increases with exposure time with lowest value of 24 for 0.1s exposures. Figure 10 shows one example of the CCD images.

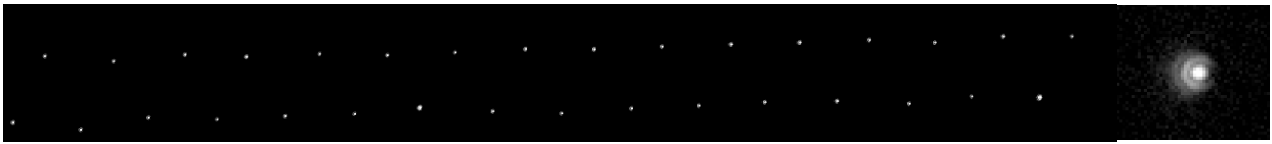

Figure 10. The FMOS fiber images taken for dome seeing measurement. The right image shows the enlarged fiber spot.

The centroids of the fibers were calculated with flat fielding by DAOPHOT. The relative movement between fibers and absolute movement of each fiber were then calculated. Finally, the RMS relative movements between fiber pairs were plotted respected to the relative distance between the two fibers. The relative distance between two fibers is calculated from the average position of each fiber over the 100 frames. In general, the files taken with shorter exposure time show larger relative image motion between fiber spots compared with the long exposure images. This fits the expectation because the higher signal to noise ratio in the long exposure images gives better centroid calculation accuracy; also the high frequency seeing effect might be averaged out. In most experiment data, the relative motion between two spots decreases as the relative distance gets shorter and approaches to some lower limit when the distance is shorter than a certain value. Figure 11 shows the result with 1s frames at different elevation angle when the dome is open.

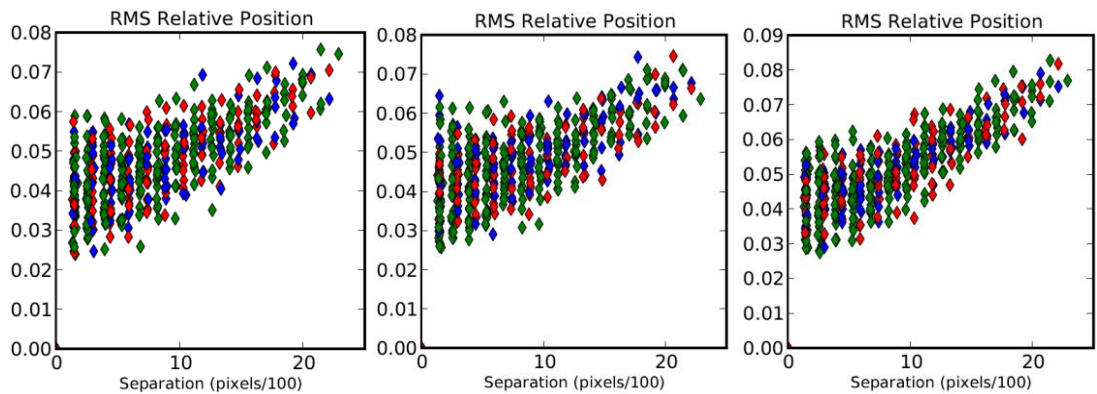

Figure 11. The RMS relative motion versus the separation of two fiber pairs when the dome was open at 30, 60 and 90 degrees of elevation angles (left to right).

The red, blue points in the plots represents the fiber pairs from the upper and lower row of the FMOS fibers. The green points are the fiber pairs from two fibers belong to different FMOS fiber rows. The three different groups of pairs show similar trend of dome seeing. This means the FMOS Echidna fiber module is quite stable. It is clearly shown that the image relative movement increases when the telescope elevation angle is higher. This is due to the larger temperature gradient through the vertical optical path. The relative motion decreases linearly with the distance between two spots. The smallest average relative motion is 0.04 pixels which is ~2.0μm at the focal plane.

In order to further understand the behavior of the turbulence, more image processing was done. We took the average fiber image positions as the reference points. A coordinate transformation between each exposure to the average frame was calculated to remove the possible global shift, tilt and most important scale change in each image. After that, the RMS image motion was calculated again. Figure 12 shows the plot for different exposure time under dome open condition with the telescope pointed at Zenith. It is clear that the distance dependence of the image motion is almost gone after the scale change. In PFS, since we have fixed fiducial fibers in the image, such scale change can be estimated and removed from the dome seeing effect. In such case, the image motion is about 0.04 pixels regardless of the fiber separation when the exposure time is longer than 1s. It should be reminded that the result shown here included the possible fiber spot center estimation error. If we removed it, the image motion is about 0.03 pixels regardless the fiber separation when the exposure time is longer than 0.5s.

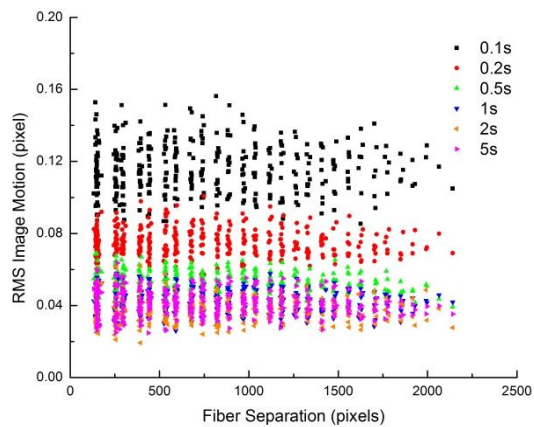

Figure 12. The RMS relative motion versus the separation of each fiber pair after the coordinate transformation when the dome was open at 90 degree elevation angle with different exposure time.

The test result shows that the possible dome seeing effect error is about 1.5μm. Combining with the other two error source mentioned, the overall centroid estimation error is about 3.4μm assuming the three factors are independent. The current design of the metrology camera meets the required 5μm error.

The dome seeing test result also suggests the exposure time for the metrology camera should be no shorter than 0.5s to keep the image motion error. If we include the readout time 0.8s and the image process time 0.6s, the overhead for each metrology camera operation is about 2s. Considering other overhead, we will be able to finish the fiber configuration in 2 minutes with 10 iterations.

## 4. SUMMARY

The design of PFS metrology camera is presented. The design of a Cassgrain metrology camera is proved to be feasible to provide the required accuracy. The optical design has been done including detailed performance and tolerance analysis. We have received the CMOS sensor in early 2014 and expect to receive the optics of PFS metrology camera by end of 2014 or early 2015. The integration and testing of PFS metrology camera will start early 2015 in Taiwan. It will be sent to Hawaii in 2016 for testing with the COBRA module and then proceed to the final integration with the COBRA operation system.

## ACKNOWLEDGEMENT

We gratefully acknowledge support from the Funding Program for World-Leading Innovative R&D on Science and Technology(FIRST) "Subaru Measurements of Images and Redshifts (SuMIRe)", CSTP, Japan for PFS project. The work in ASIAA, Taiwan is supported by the Academia Sinica of Taiwan.